\documentclass[floatfix,showkeys,amsmath,preprintnumbers,nofootinbib,onecolumn] {revtex4}
\usepackage{graphicx}
\usepackage{colordvi}
\usepackage{amssymb}
\usepackage{latexsym}

\begin{document}

\title{Evolution of Primordial Black Holes in Loop Quantum Gravity}

\author{D. Dwivedee$^a$, B. Nayak$^a$, M. Jamil$^b$ and L. P. Singh$^a$}
\affiliation{$^a${Department of Physics, Utkal University, Vanivihar,
Bhubaneswar 751004, India}\\
$^b${Center for Advanced Mathematics and Physics (CAMP),
National University of Sciences and Technology (NUST), H-12,
Islamabad, Pakistan}\\
E-mail: {debabrata@iopb.res.in, bibeka@iopb.res.in, mjamil@camp.nust.edu.pk, lambodar\_uu@yahoo.co.in}}

\begin{abstract}
In this work, we
study the evolution of Primordial Black Holes within the context of Loop Quantum
Gravity. First we calculate
 the scale factor and energy density of the universe
  for different cosmic era and then taking these as inputs we study evolution of primordial black holes.
 From our estimation it is found that accretion of
radiation does not affect evolution of primordial black holes in loop quantum gravity even though a larger number of primordial black holes may
form in early universe in comparison with Einstein's or
scalar-tensor theories.
\end{abstract}
\maketitle
\tableofcontents
\section{INTRODUCTION}
The demand for consistency between a quantum description of matter and a 
geometric description of space-time indicate 
the necessity of a complete theory of quantum gravity. This theory is 
expected to provide a  new light on singularities present in classical cosmology. 
Einstein's theory of general relativity leads to the occurrence of space-time 
singularities in a generic way. So, one may say, General Relativity is 
severely incomplete and is unable to predict what will come out of a 
singularity. One of the outstanding problems in classical Einstein cosmology 
is the Big-Bang singularity which is expected to be solved by quantum gravity. 
Loop Quantum Gravity (LQG) \cite{tam,at} is one of the best motivated theories of 
quantum gravity. LQG is a background independent, non perturbative 
approach to quantum gravity.  When Loop Quantum Gravity is applied to 
cosmology to analyse our universe, it is called Loop Quantum Cosmology(LQC) 
\cite{atp} (also see \cite{cmn} for a comprehensive review on LQC). 
In loop quantum cosmology, the non perturbative effects add a 
term of -$\rho^2/\rho_c$ to the standard Friedmann equation \cite{atp,pe}, where $\rho$ 
represents the energy density of the universe and $\rho_{c}$ is the 
critical density at which the universe is completely filled with a free massless scalar field when the scale factor reaches a minimum. The modification 
becomes important when the energy density of the universe approaches  
critical density($\rho_c$) and causes the quantum bounce. So the classical big bang is replaced by a 
quantum big bounce in such a quantum theory of gravity. Recently more and more 
researchers have paid their attention to LQC inspired by its appealing features, 
like  avoidance of various singularities \cite{mt}, inflation in loop 
quantum cosmology(LQC) \cite{xy}, large scale effect \cite{yyj}, present cosmic acceleration \cite{jamil} and so on.

        Primordial Black Holes(PBHs) are the black holes formed in the early 
universe \cite{carr}. A comparison of the cosmological density at any 
time after the Big Bang 
with the density associated with a black hole shows that PBHs would have 
mass of the order of the particle horizon mass at their formation epoch. Thus PBHs could span enermous 
mass range starting from $10^ {-5}$ g to the typical values of $10^{15}$ g. These 
black holes could be formed due to initial inhomogeneities \cite{yi,bj}, 
inflation \cite{mby,bjj}, phase transitions \cite{ma}, bubble collisions 
\cite{hmk,dp} or the decay of cosmic loops \cite{apar}. In 1974 Hawking 
discovered that the black holes emit thermal radiation due to quantum 
effects \cite{sc}. So the black holes get evaporated depending upon 
their initial  masses. Smaller the initial masses of the PBHs, quicker they evaporate. 
But the density of a black hole varies  inversely with its mass. 
So high density is required to form lighter black holes and such high 
densities are available only in the early universe. So primordial black 
holes are the only black holes whose masses could be small enough to have 
evaporated by present time. There have been speculations that PBHs could 
act as seeds for structure 
formation \cite{kjm} and could also form  a significant component of dark 
matter \cite{dda}. Since the cosmological enviornment was very hot and dense 
in the radiation dominated era, an appreciable amount of energy-matter 
from the surroundings can be absorbed by black holes. Such accretion is 
responsible for the prolongation of life time of PBHs \cite{ar,bb}.

         In this work, we study the evolution of PBH within the context of loop quantum gravity.
First, we estimate the cosmic scale factor a(t) and energy density $\rho$ (t) 
of the fluid filling the universe for different cosmic era within the 
context of loop quantum gravity. Taking these as inputs, PBH evolution 
is studied considering both the Hawking evaporation and accretion of 
radiation by the PBH. The primary aim being to compare the results so obtained with the analyses carried out earlier within the context of General Theory of Relativity and Brans-Dicke theories. 

\section{SOLUTION OF FRIEDMANN EQUATIONS}
 For a spatially flat FRW universe(k=0) with scale factor(a), two Friedmann equations in loop quantum gravity \cite{atp}, take the form :
\begin{eqnarray}
\Big(\frac{\dot{a}}{a}\Big)^2=H^2=\frac{8\pi G}{3}\rho\Big(1-\frac{\rho}{\rho_c}\Big)\\
\dot{H}=-\frac{4\pi G}{3}(\rho+p)\Big(1- \frac{2\rho}{\rho_c}\Big)
\end{eqnarray}
where H is the Hubble parameter, $\rho$ is the energy density  and P is the pressure of the fluid filling the universe.

The energy conservation equation is given by

\begin{equation}
\dot{\rho}+3H(\rho+p)=0
\end{equation}

From energy conservation equation, we get

\begin{eqnarray}
\rho\propto a^{-4}~for~t<t_e\\
\rho\propto a^{-3}~for~t>t_e 
\end{eqnarray}
where $t_e$ is the time of radiation-matter equality.

Using this solution in equation(1), one gets the temporal behaviour of the scale factor $a(t)$ as shown below.

$\textbf{For radiation dominated era $(t<t_e)$}$ :
\begin{equation}
a(t)=\Big[\frac{\rho_0 {a_0}^3 a_e}{\rho_c}+\Big{\{}2{\rho_0}^{1/2}{a_0}^{3/2}{a_e}^{1/2}\sqrt{\frac{8\pi G}{3}}(t-t_e)+{\Big({a_e}^4-\frac{\rho_0{a_0}^3a_e}{\rho_c}\Big)}^{1/2}\Big{\}}^2\Big]^{1/4}
\end{equation}

where the subscript 0 indicates the present value of any parameter and $a_e =a(t_e)$
and $\rho_c$ represents the critical value of energy density of the universe given by $\rho_c=\frac{\sqrt3}{16\pi^2 \gamma^3}\rho_{pl}$ with $\gamma =\frac{ln2}{\pi \sqrt3}$ is the dimensionless Barbero-Immirzi parameter \cite{amk} and  $\rho_{pl}$ is the energy density of universe in Planck time. 

$\textbf{For matter dominated era  $(t>t_e)$}$ :
\begin{equation}
a(t)=\Big[\frac{\rho_0 {a_0}^3}{\rho_c}+\Big{\{}\frac{3}{2}\rho_0^{1/2}a_0^{3/2}\sqrt{\frac{8\pi G}{3}}(t-t_0)+{\Big(a_0^3-\frac{\rho_0}{\rho_c}a_0^3\Big)}^{1/2}\Big{\}}^2\Big]^{1/3}
\end{equation}

Using equations (4) and (6), we get

$\underline{For~t<t_e}$

\begin{eqnarray}
\rho(t)=\rho_0\Big[\frac{\rho_0}{\rho_c}+\Big{\{}2\sqrt{\frac{8\pi G}{3}}{\rho_0}^{1/2}(t-t_e)+\frac{3}{2}\sqrt{\frac{8\pi G}{3}}{\rho_0}^{1/2}(t_e-t_0)  \nonumber \\ +\Big(1-\frac{\rho_0}{\rho_c}\Big)^{1/2}\Big{\}}^2\Big]^{-1}
\end{eqnarray}

Using equations (5) and (7), we get

$\underline{For~t>t_e}$

\begin{equation}
\rho(t)=\rho_0\Big[\frac{\rho_0}{\rho_c}+\Big{\{}\frac{3}{2}\sqrt{\frac{8\pi G}{3}}~\rho_0^{1/2}~(t~-~t_0)~+~\Big(1~-~\frac{\rho_0}{\rho_c}\Big)^{1/2}~\Big{\}}^{2}\Big]^{-1}
\end{equation}

\section{ACCRETION OF RADIATION }

When a PBH evolves through radiation dominated era, it can also accrete radiation from the surrounding. The accretion of radiation leads to increase of its mass with the rate given by

\begin{equation}
{\dot{M}}_{acc}=4\pi f{r_{BH}}^2\rho_r
\end{equation}
where $\rho_r$ is the radiation energy density of the surrounding of the 
black hole,  $r_{BH}$ is the black hole radius and $f$ is the accretion 
efficiency. The value of the accretion efficiency $f$ depends on the 
complex physical processes such as the mean free path of the particles 
comprising the radiation surrounding PBHs. Any peculiar velocity of the PBH 
with respect to the cosmic frame could increase the value of $f$ \cite{apr,rda}.
 Since the precise value of $f$ is unknown, it is customary \cite{ds} 
to take the accretion rate to be proportional to the product of the surface 
area of the PBH and the energy density of radiation with $f\sim$ O(1).

After substituting the expressions for $r_{BH}$ = 2GM and $\rho_r$ given by equation(8) in equation (10) , we get

\begin{eqnarray}
\dot{M}_{acc}= 16\pi f G^2 M^2 \rho_0\Big[\frac{\rho_0}{\rho_c} + \Big{\{}2\sqrt{\frac{8\pi G}{3}} \rho_0^{1/2}(t-t_e)+\frac{3}{2}\sqrt{\frac{8\pi G}{3}} \rho_0^{1/2} (t_e-t_0)+ \nonumber \\ (1-\frac{\rho_0}{\rho_c})^{1/2}\Big{\}}^2 \Big]^{-1}
\end{eqnarray}

Solving equation(11), one can find 

\begin{eqnarray}
M(t)=\Big{\{}{M_i}^{-1}-8\pi f G^2 {\rho_c}^{1/2} \sqrt{\frac{3}{8\pi G}} tan^{-1}[ {\frac{\rho_c}{\rho_0}} Z(t)- 1]^{1/2}+8\pi f G^2 {\rho_c}^{1/2}\sqrt{\frac{3}{8\pi G}} \nonumber \\ tan^{-1}[{\frac{\rho_c}{\rho_0}} Z(t_i)-1]^{1/2}\Big{\}}^{-1}
\end{eqnarray}

Again using horizon mass as initial mass of PBH i.e; $M_{i}$ = $M_{H}(t_{i})$=$G^{-1}t_{i}$, we get

\begin{equation}
M(t)=M_i[1-8\pi f G^{1/2}{\rho_c}^{1/2}{t_i}\sqrt{\frac{3}{8\pi}}\{ tan^{-1}[ \frac{\rho_c}{\rho_0} Z(t)-1]^{1/2}-tan^{-1}[\frac{\rho_c}{\rho_0}Z(t_i)-1]^{1/2}\}]^{-1}
\end{equation}

where $Z(t)=\frac{\rho_0}{\rho_c}+\{ 2\sqrt\frac{8\pi G}{3}{\rho_0}^{1/2}(t-t_e)+\frac{3}{2}\sqrt{\frac{8\pi G}{3}}{\rho_0}^{1/2}(t_e-t_0)+(1-\frac{\rho_0}{\rho_c})^{1/2}\}^2$.

\section{EVAPORATION OF PBH}
As is well known black holes can also loose mass through Hawking evaporation. The rate at which the PBH mass decreases due to evaporation is given by

\begin{equation}
{\dot{M}}_{evap}= -4\pi r_{BH}^2 a_{H}T_{BH}^4
\end{equation}

where   $a_{H}\sim$  is the Blackbody Constant and $T_{BH}\sim$ is the Hawking Temperature = $\frac{1}{8\pi G M}$

Now equation(14) becomes
\begin{equation}
\dot{M}_{evap}=-\frac{a_H}{256\pi^3}~ \frac{1}{G^2 M^2}
\end{equation}

Integrating the above equation, we get

\begin{equation}
M~=~\Big[{M_i}^3~+~3\alpha(t_i-t)\Big]^{1/3}
\end{equation} 

where $\alpha~=~ \frac{a_H}{256 \pi^3}~\frac{1}{G^2}$\\

and $M_{i}$ is the the black hole mass at its formation time $t_{i}$.\\
We rewrite equation (16) as 

\begin{equation}
M(t)~=~M_i\Big[1~+~\frac{3\alpha}{M_{i}^3}(t_i~-~t)\Big]^{1/3}
\end{equation}             

\section{PBH DYNAMICS IN DIFFERENT ERA}

Primordial Black Holes, as discussed earlier, are formed only in radiation dominated era. We now study PBHs so formed in two categories:

(i) PBHs evaporating in radiation dominated era $(t_{evap}<t_e)$\\
(ii)PBHs evaporating in matter dominated era $(t_{evap}>t_e)$

\underline{CASE-1} ($t_{evap}<t_e$)

If we consider both evaporation and accretion simultaneously, then the rate at which primordial black hole mass changes is given by

\begin{eqnarray}
\dot{M}_{PBH} ~=~16\pi fG^2 M^2\rho_0\Big[\frac{\rho_0}{\rho_c}~+~\Big{\{}2\sqrt\frac{8\pi G}{3} {\rho_0}^{1/2}(t~-~t_e)~ \nonumber \\ +~\frac{3}{2}\sqrt\frac{8 \pi G}{3}{\rho_0}^{1/2}(t_e~-~t_0)~  +~\Big(1~-~\frac{\rho_0}{\rho_c}\Big)^{1/2}\Big{\}}^2\Big]^{-1}~-~\frac{a_{H}}{256\pi^3}\frac{1}{G^2 M^2}
\end{eqnarray}

Solving above equation numerically, we construct the following table for PBHs which are evaporating in radiation dominated era. In our calculation, we have used
 $\rho_0 = 1.1 \times 10^{-29}g-cm^{-3}$,
$\rho_c = 5.317 \times 10^{94}g-cm^{-3} $,
$G = 6.673 \times 10^{-8}dyne-cm^2/g^2$,
$t_e = 10^{11}s$,
$t_0 = 4.42 \times 10^{17}s$ and 
$M_e = 10^{49}g$.
 
\begin{table}[h]
\begin{tabular}[c]{|c|c|c|c|}
\hline
$t_i$ & $M_i$ & \multicolumn{2}{|c|}{$t_{evap}$} \\
 &  & $f=0$ & $f=1$\\
\hline
$10^{-32}s$ & $10^{6}g$ & $3.333\times10^{-11}s$ & $3.333\times10^{-11}s$\\
\hline
$10^{-30}s$ & $10^{8}g$ & $3.333\times10^{-5}s$ & $3.333\times10^{-5}s$\\
\hline
$10^{-28}s$ & $10^{10}g$ & $3.333\times10^{1}s$ & $3.333\times10^{1}s$\\
\hline
$10^{-26}s$ & $10^{12}g$ & $3.333\times10^{7}s$ & $3.333\times10^{7}s$\\
\hline
\end{tabular}
\caption{Display of formation times and initial masses of the PBHs evaporating in radiation dominated era} 
\end{table}

It is clear from Table-1 that with increase in initial mass, evaporating time increases. However radiation accretion, surprisingly, seems to have little effect on evolution of PBH unlike the results obtained in theories of Einstein or scalar-tensor type. This is also shown in Figure-1.

\begin{figure}[h]
\centering
\includegraphics{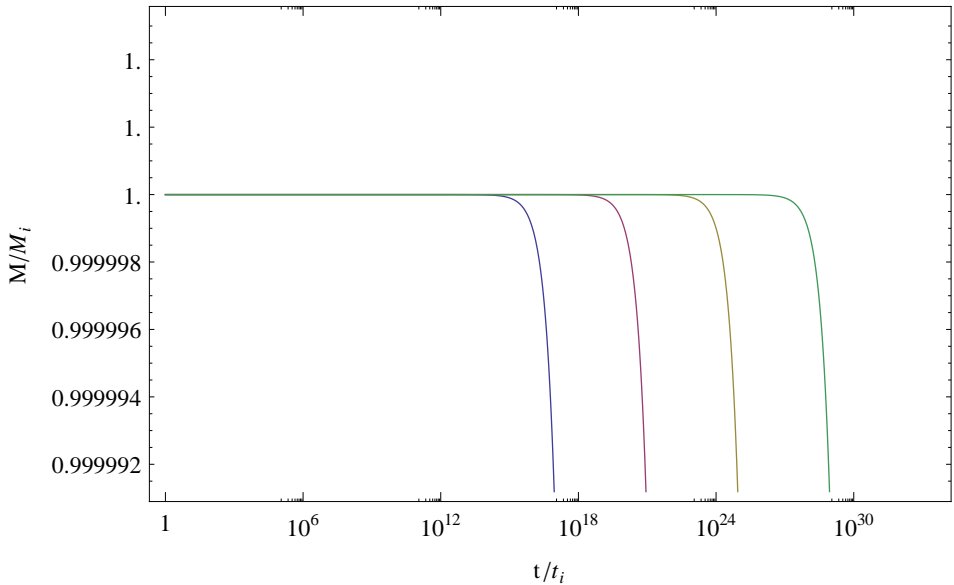}
\caption{Evaporation of PBHs for different initial masses (i.e. $10^6g$, $10^8g$, $10^{10}g$ and $10^{12}g$) are shown in the Figure where axes are taken in logarithmic scale}
\label{fig1}
\end{figure}

$~~$\\

\underline{CASE-II} ($t_{evap}>t_e$)

Since there is insignificant accretion of radiation in matter dominated era,  the first term in the combined equation(18) for variation of $M_{PBH}$ with time should be integrated only upto $t_{e}$. Basing upon the numerical solution we construct Table-2 for the PBHs evaporating by the present time.  
\begin{table}[h]
\begin{tabular}[c]{|c|c|c|}
\hline
\multicolumn{3}{|c|}{${t_{evap} =t_0 =4.42\times10^{17}s}$}\\
\hline
$f$ & $t_i$ & $M_i$ \\
\hline
0 & $2.3669\times10^{-23}s$ & $2.3669\times10^{15}g$\\
\hline
0.25 & $2.3669\times10^{-23}s$ & $2.3669\times10^{15}g$\\
\hline
0.5 & $2.3669\times10^{-23}s$ & $2.3669\times10^{15}g$\\
\hline
1.0 & $2.3669\times10^{-23}s$ & $2.3669\times10^{15}g$\\
\hline
\end{tabular}
\caption{Display of formation times of PBHs which are evaporating now for several accretion efficiencies}
\end{table}

It is clear from Table-2 that PBH evaporation is again not affected by radiation accretion efficiency.
\section{CONSTRAINTS ON PBH}
The fraction of the Universe mass going into PBHs at time t is [9] 
\begin{equation}
\beta(t)=\Big[\frac{\Omega_{PBH}(t)}{\Omega_{R}}\Big] (1+z)^{-1}  
\end{equation} 

where $\Omega_{PBH}(t)$ is the density parameter associated with PBHs formed at time t, z is the redshift associated with time t and $\Omega_{R}$ is the microwave background density having value $10^{-4}$.

Substituting the value of $\Omega_{R}$ in the above equation, we get
\begin{equation}
\beta(t)= (1+z)^{-1} \Omega_{PBH}(t) \times 10^4 
\end{equation}
For $ t<t_e$, redshift definition implies 
\begin{equation}
(1+z)^{-1} = \frac{a(t)}{a(t_0)}
            =\frac{a(t)}{a(t_e)} \frac{a(t_e)}{a(t_0)}
\end{equation}

But here
\begin{eqnarray}  
\frac{a(t)}{a(t_e)} = \Big[\frac{\rho_0}{\rho_c} \frac{a_0^3}{a_e^3} + \Big\{ 2\rho_0^{1/2} \frac{a_0^{3/2}}{a_e^{3/2}} \sqrt\frac{8\pi G}{3} (t-t_e) + \Big(1 -{\frac{\rho_0}{\rho_c}} {\frac{a_0^3}{a_e^3}}\Big)^{1/2}\Big{\}}^2 \Big ]^{1/4}  
\end{eqnarray}
and
\begin{eqnarray}
\frac{a(t_e)}{a(t_0)} = \Big[ \frac{\rho_0}{\rho_c} + \Big \{ \frac{3}{2} \rho_o^{1/2}\sqrt{\frac{8\pi G}{3}}(t_e-t_0) + \Big(1 - \frac{\rho_0}{\rho_c}\Big)^{1/2} \Big{\}}^2 \Big]^{1/3}  
\end{eqnarray} 

Using above numerical values of different quantities in equation(23), we get
\begin{equation}
\frac{a(t_e)}{a(t_0)} = 0.746 
\end{equation}

Using equations (22) and (24) in equation (21), we get

\begin{equation}
(1+z)^{-1}= \Big[\frac{\rho_0}{\rho_c} \frac{a_0^3}{a_e^3} + \Big\{ 2\rho_0^{1/2} \frac{a_0^{3/2}}{a_e^{3/2}} \sqrt\frac{8\pi G}{3} (t-t_e) + \Big(1 -{\frac{\rho_0}{\rho_c}} {\frac{a_0^3}{a_e^3}}\Big)^{1/2}\Big{\}}^2 \Big ]^{1/4} \times 0.746 
\end{equation}

Substituting equation(25) in equation (20), we get
  
\begin{equation}
\beta(t) = \Big[\frac{\rho_0}{\rho_c} \frac{a_0^3}{a_e^3} + \Big\{ 2\rho_0^{1/2} \frac{a_0^{3/2}}{a_e^{3/2}} \sqrt\frac{8\pi G}{3} (t-t_e) + \Big(1 -{\frac{\rho_0}{\rho_c}} {\frac{a_0^3}{a_e^3}}\Big)^{1/2}\Big{\}}^2 \Big ]^{1/4} \times 0.746 \times \Omega_{PBH}(t) \times 10^4
\end{equation} 

Using $M = G^{-1}t$, we can write equation (26) to represent the fraction of the Universe going into PBHs' as a function of mass M as

\begin{equation}
\beta(M) = \Big[\frac{\rho_0}{\rho_c} \frac{a_0^3}{a_e^3} + \Big\{ 2\rho_0^{1/2} \frac{a_0^{3/2}}{a_e^{3/2}} \sqrt\frac{8\pi G}{3}G (M-M_e) + \Big(1 -{\frac{\rho_0}{\rho_c}} {\frac{a_0^3}{a_e^3}}\Big)^{1/2}\Big{\}}^2 \Big ]^{1/4} \times 0.746 \times \Omega_{PBH}(M) \times 10^4
\end{equation}
Observations of the cosmological deceleration parameter imply that 
$\Omega_{BH}(M) < 1$ over all mass ranges for which PBHs have not evaporated 
yet. But presently evaporating $ PBHs(M_*)$ generate a $ \gamma$- ray 
background whose most of the energy is appearing at around 100 Mev \cite{bjk}.
 If the fraction of the emitted energy which goes into photons is 
$\epsilon_{\gamma}$, then the density of the radiation at this energy 
is expected to be $\Omega_{\gamma} = \epsilon_{\gamma} \Omega_{PBH}(M_{*})$. 
Since $\epsilon_{\gamma}\sim  0.1$ \cite{upase} and the observed $\gamma$ -ray 
background density around 100 Mev is $\Omega_{\gamma} \sim 10^{-9}$, 
we get $\Omega_{PBH} < 10^{-8}$.

With the use of all these parameters, Eqn (27) leads to an upper bound
\begin{equation}
\beta(M_*) < 0.746 \times 10^{-4} 
\end{equation} 
Here for the presently evaporating PBHs the upper limit is much greater than previous results \cite{bb,idn,jb} 
obtained by assuming Einstein's theory or Brans Dicke theory as the theory 
of gravity. But from our calculation, we found that the formation time of 
presently evaporating PBHs is nearly same in all theories. This higher upper bound implies that in LQC a much larger number of PBHs would form 
in early Universe in comparison with standard cosmology and scalar-tensor 
theories.  

\section{CONCLUSION}

We  have studied PBH evolution in loop quantum gravity. We have estimated the cosmic scale factor a(t) and energy density $\rho$ (t) of the 
universe for both radiation dominated era and matter dominated era. Both 
expressions for a(t) and $\rho$(t) are different from those in standard 
cosmology. In the limit $\rho_{c} \to \infty$ these expressions go over to those of standard cosmology since standard cosmology envisages nearly zero size for the universe at the time of creation. Using these results as inputs we have studied evolution of PBHs using both accretion of radiation and evaporation. We find accretion of radiation has no effect on PBH 
evaporation in the present formalism. From numerical calculation it is found that the PBHs 
created before $1.443\times 10^{-25}$ s could evaporate completely in 
radiation dominated era and the accretion efficiency does not affect the 
evaporation of individual PBHs formed at different times in this era. Further, 
we found that the upper bound on initial PBH mass fraction is much greater 
than all previous analyses but the formation time of presently evaporating 
PBHs is nearly the same. The greater upperbound implies that a large number of PBHs could possibly form in early universe within the context of LQG in comparison with standard 
cosmology and scalar-tensor theories.

\section*{ACKNOWLEDGEMENT}
One of our author B. Nayak would like to thank the Council of Scientific and Industrial Research, Government of India, for the award of SRF, F. No. 09/173(0125)/2007-EMR-I. We are thankful to Institute of Physics, Bhubaneswar, India, for providing the library and computational facilities.


\begin{thebibliography}{99} 
\bibitem{tam}
T. Thiemann, Modern Canonical Quantum General relativity, 
Cambridge University, Press  2007; 
A. Ashtekar and J.Lewandowski, 
\textit{Background independent quantum gravity: A Status report},  
Class.Quant.Grav.$\textbf{21}$ (2004) R53 [gr-qc/0404018]; 
M. Han, Y. Ma, W. Huang, 
\textit{Fundamental structure of loop quantum gravity}, 
Int. J. Mod. Phys. D $\textbf{16}$ (2007) 1397 [gr-qc/0509064].

\bibitem{at}
C. Rovelli, $\textit{Quantum gravity}$ 
(Cambridge University Press, Cambridge, England, 2004); 
T. Thiemann, \textit{Modern canonical quantum general relativity},[gr-qc/0110034].

\bibitem{atp}
A. Ashtekar, T. Pawlowski and P. Singh, 
\textit{Quantum nature of the big bang}, 
Phys. Rev. Lett. $\textbf{96}$ (2006) 141301 [gr-qc/0602086]; 
\textit{Quantum Nature of the Big Bang: Improved dynamics}, 
Phys.Rev. D$\textbf{74}$ (2006) 084003 [gr-qc/0607039].

\bibitem{cmn} K. Banerjee, G. Calcagni, M. M-Benito, 
\textit{Introduction to quantum cosmology} [arXiv:1109.6801].

\bibitem{pe}
P. Singh, 
\textit{Loop cosmological dynamics and dualities with Randall-Sundrum braneworlds},
Phys. Rev. D$\textbf{73}$ (2006) 063508 [gr-qc/0603043]; 
E.J. Copeland, J.E. Lidsey and S.Mizuno, 
\textit{Correspondence between loop-inspired and braneworld cosmology},
Phys. Rev. D$\textbf{73}$ (2006) 043503 [gr-qc/0510022].

\bibitem{mt}
M. Sami, P. Singh and S. Tsujikawa, 
\textit{Avoidance of future singularities in loop quantum cosmology},
Phys. Rev. D$\textbf{74}$ (2006) 043514 [gr-qc/0605113]; 
T. Cailleteau, A. Cardoso, K. Vandersloot and D. Wands, 
\textit{Singularities in loop quantum cosmology},
Phys. Rev. Lett.$\textbf{101}$ (2008) 251302 [arXiv:0808.0190]. 

\bibitem{xy}
X. Zhang and Y. ling, 
\textit{Inflationary universe in loop quantum cosmology},
JCAP $\textbf{08}$ (2007) 012 [arXiv:0705.2656]; 
 M. Bojowald,
G. Calcagni, S. Tsujikawa [arXiv:1101.5391];
M. Bojowald, G. Calcagni, 
\textit{Inflationary observables in loop quantum cosmology}, 
JCAP $\textbf{03}$ (2011) 032 [arXiv:1011.2779];  
M. Bojowald,
\textit{Consistent Loop Quantum Cosmology},
 Class. Quant. Grav. $\textbf{26}$ (2009) 075020 [arXiv:0811.4129];  
M. Bojowald, 
\textit{The Early universe in loop quantum cosmology},
J. Phys. Conf. Ser. $\textbf{24}$ (2005) 77 [gr-qc/0503020]. 

\bibitem{yyj}
Y. Ding, Y. Ma and J. yang, 
\textit{Effective Scenario of Loop Quantum Cosmology},
Phys. Rev. Lett.$\textbf{102}$ (2009) 051301 [arXiv:0808.0990];
M. Bojowald, H. Hernandez, M. Kagan, P. Singh, A.
Skirzewski, 
\textit{Formation and Evolution of Structure in Loop Cosmology}, 
Phys. Rev. Lett. $\textbf{98}$ (2007) 031301 [astro-ph/0611685]; 
M. Bojowald,
\textit{Large scale effective theory for cosmological bounces},
 Phys. Rev. D $\textbf{74}$ (2007) 081301 [gr-qc/0608100]; 
M. Bojowald, 
\textit{Loop quantum cosmology},
Living Rev. Rel. $\textbf{8}$ (2005) 11 [gr-qc/0601085].

\bibitem{jamil}
 M. Jamil, D. Momeni, M. A. Rashid,
\textit{ Notes on dark energy interacting with dark matter and unparticle in loop quantum cosmology},
 Eur. Phys. J. C $\textbf{71}$ (2011) 1711 [arXiv:1107.1558];
M. Jamil and U. Debnath,
\textit{Interacting modified Chaplygin gas in loop quantum cosmology},
 Astrophys. Space Sci. $\textbf{333}$ (2011) 3 [arXiv:1102.2758]; 
K Xiao, J-Y Zhu, 
\textit{Dynamical behavior of interacting dark energy in loop quantum cosmology},
Int. J. Mod. Phys. A $\textbf{25}$ (2010) 4993 [arXiv:1006.5377]; 
M. Jamil, M. U. Farooq, 
\textit{Interacting holographic dark energy with entropy corrections},
JCAP $\textbf{03}$ (2010) 001 [arXiv:1002.1434]; 
S. Chen, B. Wang, J. Jing,
\textit{Dynamics of interacting dark energy model in Einstein and Loop Quantum Cosmology}
 Phys. Rev. D $\textbf{78}$ (2008) 123503 [arXiv:0808.3482]; 
H.M. Sadjadi, M. Jamil,
\textit{Cosmic accelerated expansion and the entropy corrected holographic dark energy},
Gen. Rel. Grav. $\textbf{43}$ (2011) 1759 [arXiv:1005.1483];
X. Fu, H. Yu, P. Wu,
\textit{Dynamics of interacting phantom scalar field dark energy in Loop Quantum Cosmology},
 Phys. Rev. D $\textbf{78}$ (2008) 063001 [arXiv:0808.1382]; 
P. Wu, S. N. Zhang, 
\textit{Cosmological evolution of interacting phantom (quintessence) model in Loop Quantum Gravity},
JCAP $\textbf{06}$ (2008) 007 [arXiv:0805.2255]; 
G. Cognola, E. Elizalde, S. Nojiri,
S. D. Odintsov, S. Zerbini,
\textit{One-loop f(R) gravity in de Sitter universe},
 JCAP $\textbf{02}$ (2005) 010 [hep-th/0501096].

\bibitem{carr} 
B.J. Carr, 
\textit{Primordial black holes as a probe of cosmology and high energy physics}, Lect. Notes Phys. $\textbf{631}$ (2003) 301 [astro-ph/0310838]; 
M. Jamil and A. Qadir, 
\textit{Primordial Black Holes in Phantom Cosmology},
Gen. Rel. Grav. $\textbf{43}$ (2011) 1069 [arXiv:0908.0444]; 
M. Jamil, \textit{Black Holes in an Accelerated Universe}, 
Lambert Academic Publishing, 2011 ; 
M. Jamil,
\textit{Black Holes in Accelerated Universe},
 Int. J. Theor. Phys. $\textbf{49}$ (2010) 1706 [arXiv:0806.1320]. 

\bibitem{yi}
Ya. B. Zeldovich and I. Novikov, 
\textit{The hypothesis of cores retarded during expansion and the hot cosmological model}, 
Sov. Astron. $\textbf{10}$ (1967) 602.

\bibitem{bj}
B. J. Carr, 
\textit{The Primordial black hole mass spectrum}, 
Astrophys. J. $\textbf{201}$ (1975) 1.

\bibitem{mby}
M. Y. Kholpov, B. A. Malomed and  Ya. B. Zeldovich, 
\textit{Gravitational instability of scalar fields and formation  of primordial black holes}, 
Mon. Not. R. Astron. Soc. $\textbf{215}$ (1985) 575.

\bibitem{bjj}
B. J. Carr, J. Gilbert and J. Lidsey, \textit{Black hole relics and inflation: Limits on blue perturbation spectra}, 
Phys. Rev. D$\textbf{50}$ (1994) 4853.

\bibitem{ma}
M. Y. Kholpov and A. G. Polnarev, \textit{Primordial Black Holes As A Cosmological Test of Grand Unification},
Phys. Lett.$\textbf{B97}$ (1980) 383. 

\bibitem{hmk}
H. Kodma, M. Sasaki and K. Sato, \textit{Abundance of Primordial Holes Produced By Cosmological First-order Phase Transition}, Prog. 
Theor. Phys. $\textbf{68 }$ (1982) 1979.

\bibitem{dp}
D. La and P. J. Steinhardt, 
\textit{Extended Inflationary Cosmology},
Phys. Rev. Lett $\textbf{62 }$ (1989) 376.

\bibitem{apar}
A. Polnarev and R. Zemboricz, \textit{Formation of primordial black holes by cosmic strings},
Phys. Rev. D $\textbf{43}$ (1991) 1106.

\bibitem{sc}
S. W. Hawking, \textit{Particle Creation by Black Holes},
Commun. Math. Phys. $\textbf{43}$ (1975) 199.

\bibitem{kjm}
K. J. Mack, J. P. Ostriker and M. Ricotti, 
\textit{Growth of structure seeded by primordial black holes},
Astrophys. J. $\textbf{665}$ (2007) 1277 [astro-ph/0608642].

\bibitem{dda}
D. Blais, C. Kiefer, D. Polarski, 
\textit{Can primordial black holes be a significant part of dark matter},
Phys. Lett. B $\textbf{535}$ (2002) 11 [astro-ph/0203520]; 
D. Blais, T. Bringmann, C. Kiefer, D. Polarski,
\textit{ Accurate results for primordial black holes from spectra with a distinguished scale}, 
Phys. Rev. D $\textbf{67}$ (2003) 024024 [astro-ph/0206262];  

\bibitem{ar}
A. S. Majumdar, \textit{Domination of black hole accretion in brane cosmology},
Phys. Rev. Lett. $\textbf{90}$ (2003) 031303 [astro-ph/0208048]; 
A. S. Majumdar and N. Mukharjee, \textit{Braneworld black holes in cosmology and astrophysics},
Int. J. Mod. Phys. D $\textbf{14}$ (2005) 1095 [astro-ph/0503473]; 
A. S. Majumdar, D. Gangopadhyay and L. P. Singh, \textit{Evolution of primordial black holes in Jordan-Brans-Dicke cosmology}, 
Mon. Not. R. Astron. Soc. $\textbf{385}$ (2008) 1467 [arXiv:0709.3193].

\bibitem{bb}
B. Nayak, L. P. Singh and A. S. Majumdar,
\textit{Effect of accretion on primordial black holes in Brans-Dicke theory}, 
Phys. Rev. D $\textbf{80}$ (2009) 023529 [arXiv:0902.4553];
B. Nayak, A. S. Majumdar and L. P. Singh, 
\textit{ Astrophysical constraints on primordial black holes in Brans-Dicke theory}, 
JCAP $\textbf{08}$ (2010) 039 [arXiv:1002.3472];
B. Nayak and L. P. Singh, 
\textit{Note on nonstationarity and accretion of Primordial Black Holes in Brans-Dicke theory}, 
Phys. Rev. D $\textbf{82}$ (2010) 127301 [arXiv:1005.1529]; 
B. Nayak and L. P. Singh, 
\textit{ Accretion, Primordial Black Holes and Standard Cosmology}, 
Pramana $\textbf{76}$ (2011) 173 [arXiv:0905.3243]; 
B. Nayak and M. Jamil, \textit{ Effect of Vacuum Energy on Evolution of Primordial Black Holes in Einstein Gravity} [arXiv:1107.2025].

\bibitem{amk}
A. Ashtekar, J. Baez, A. CCorichi and K. Krasnov, 
\textit{Quantum geometry and black hole entropy},
Phys. Rev. Lett. $\textbf{80}$ (1998) 904 [gr-qc/9710007]; 
M. Domagala and J. Lewandowski,
\textit{Black hole entropy from quantum geometry},  
Class. Quant. Grav.$\textbf{21}$ (2004) 5233 [gr-qc/0407051]; 
K. A. Meissner, 
\textit{Black hole entropy in loop quantum gravity},
Class. Quant. Grav. $\textbf{21}$ (2004) 5245 [gr-qc/0407052].

\bibitem{apr}
A. S. Majumdar, P. Das Gupta and R. P. Saxena,
\textit{Baryogenesis from black hole evaporation}, 
Int. J. Mod. Phys. D $\textbf{4}$ (1995) 517. 

\bibitem{rda}
R. Guedens, D. Clancy and A. R. Liddle, \textit{Primordial black holes in braneworld cosmologies:Accretion after formation},
Phys. Rev. D $\textbf{66}$ (2002) 083509.

\bibitem{ds}
D.N. Page and S. W. Hawking, \textit{Gamma rays from primordial black holes},
Astrophys. J. $\textbf{206}$ (1976) 1.

\bibitem{bjk}
B. J. Carr, 
\textit{Quark jets from evaporating black holes}, 
Astron. Astrophys. Trans. $\textbf{5}$ (1994) 43.

\bibitem{upase}
D. N. Page, \textit{Particle Emission Rates from a Black Hole: Massless Particles from an Uncharged,Nonrotating Hole}, Phys. Rev. D $\textbf{13}$ (1976) 198.

\bibitem{idn}
I. D. Novikov et al., 
\textit{Primordial black holes}, 
Astron. Astrophys. J. $\textbf{80}$ (1979) 104.

\bibitem{jb}
J. MacGibbon and B. J. carr, 
\textit{ Cosmic rays from primordial black holes}, 
Astrophys. J. $\textbf{371}$ (1991) 447.

\end{thebibliography}
\end{document}